\documentclass[prd,aps,secnumarabic,superscriptaddress,floatfix,preprintnumbers]{revtex4}
\usepackage{dcolumn}
\usepackage{amsmath}
\usepackage[dvips]{graphicx,color}
\usepackage{bm}
\def\Sc{\mbox{\rule[-5pt]{0pt}{16pt}}}
\def\ds{\displaystyle}
\def\Sb{\mbox{\rule{0pt}{11pt}}}
\def\Sw{\mbox{\rule{24pt}{0pt}}}
\def\al{\alpha}

\def\ga{\gamma}
\def\de{\delta}

\def\om{\omega}
\def\th{\theta}
\def\Ga{\Gamma}

\def\Dot{\!\cdot\!}

\begin{document}
\preprint{NSF-KITP-09-08}
\vspace{12pt}
\title{Potential model calculations and predictions for $\bm c\bar{
\bm s}$ quarkonia}
\author{Stanley F. Radford}\email{sradford@brockport.edu}
\affiliation{Department of Physics, The College at Brockport, State
University of New York, Brockport, NY 14420 \\ and \\
Kavli Institute for Theoretical Physics, Santa Barbara, CA 93106}

\author{Wayne W. Repko}\email{repko@pa.msu.edu}
\author{Michael J. Saelim}\email{mike.saelim@gmail.com}
\affiliation{Department of Physics and Astronomy, Michigan State University, East Lansing, MI 48824} 
\date{\today}
\begin{abstract}
We investigate the spectroscopy and decays of the charm-strange quarkonium system in a potential model consisting of a relativistic kinetic energy term, a linear confining term including its scalar and vector relativistic corrections and the complete perturbative one-loop quantum chromodynamic short distance potential. The masses and wave functions of the various states are obtained using a variational technique, which are then used in a perturbative
treatment of the potential to find the mass spectrum of the $c\bar{s}$ system and radiative decay widths. Our results compare well with the available data for the spectrum of $D_s$ states. We include a discussion of the effect of mixing and an investigation of the Lorentz nature of the confining potential.
\end{abstract}
\maketitle

\section{INTRODUCTION}

Recently we reported on a study of the charmonium and upsilon
systems in a semi-relativistic model which includes all $v^2/c^2$
and one-loop QCD corrections for the interaction of a quark and
antiquark of equal mass\cite{rr1}. This semi-relativistic potential
model successfully describes the spectra and leptonic and radiative
decays of those systems. We have now extended this modelling
approach to systems in which the quark and antiquark have different
masses.

Interest in the modelling of light-heavy quarkonia is over 25 years
old \cite{grr1}. A variety of modelling approaches have been employed
with varying success \cite{gj1,godk,god,cj,beh,ls}.  Renewed and continuing
interest in the modelling of $c\bar{s}$ quarkonia is fueled by, in
particular, the recent discovery of the $2^3S_1$ state \cite{ab} as
well as ongoing efforts to determine the masses and decays of the
$D_s$ mesons \cite{expts}.

We have revised and extended the approach of our earlier papers in
order to investigate the spectroscopy and decays of the $D_s$
system, as well as to discuss other questions of modelling interest.
In addition, we investigate the the scalar/vector mixture of the
phenomenological confining potential

In the next Section, we describe the potential model in some detail.
This is followed, in Section \ref{III}, by an outline of our
calculational approach. In Section \ref{IV}, we present our results
for the $D_s$ system, and then give some conclusions in Section
\ref{V}. The conventions we use for our treatment of the mixing of the $J=1$ $p-$states are given in the Appendix.
\section{SEMI-RELATIVISTIC MODEL \label{II}}

In our analysis, we use a semi-relativistic Hamiltonian of the
general form
\begin{eqnarray}
H&=&\sqrt{\vec{p}^{\,2}+m_1^2}+\sqrt{\vec{p}^{\,2}+m_2^2}+Ar-\frac{4\al_S}{3r}\left[1-\frac{3\al_S}{2\pi}+\frac{\al_S}{6\pi}(33-2n_f) \left(\ln(\mu r)+\ga_E\right)\right] +V_L+V_S \\
&=&H_0+V_L+V_S\,,
\end{eqnarray}
where $m_1$ and $m_2$ are the quark masses, $\mu$ is the
renormalization scale, $n_f$ is the effective number of light quark
flavors and $\gamma_E$ is Euler's constant. $V_L$ contains the
$v^2/c^2$ corrections to the linear confining potential
\begin{eqnarray} \label{VL}
V_{L}&=&-(1-f_V)\frac{A}{4r}\left[(\frac{1}{m_1^2}+\frac{1}{m_2^2})\vec{L}\Dot\vec{S}+(\frac{1}{m_1^2}-\frac{1}{m_2^2})\vec{L}\Dot(\vec{S_1}-\vec{S_2})\right]\nonumber \\
& &
+f_V\frac{A}{4r}\left[\frac{1}{m_1^2}+\frac{1}{m_2^2}+\frac{16}{3m_1m_2}\vec{S_1}\Dot\vec{S_2}+
(\frac{1}{m_1^2}+\frac{1}{m_2^2}+\frac{4}{m_1m_2})\vec{L}\Dot\vec{S}+(\frac{1}{m_1^2}-\frac{1}{m_2^2})\vec{L}\Dot(\vec{S_1}-\vec{S_2})\right.
\nonumber \\
&
&\left.+\frac{4}{3m_1m_2}(3\vec{S_1}\Dot\hat{r}\vec{S_2}\Dot\hat{r}-\vec{S_1}\Dot\vec{S_2})\right],
\end{eqnarray}
where $A$ is the linear coupling strength. The first line in Eq.(3)
is the contribution from scalar exchange while the second and third
lines give the contribution from vector exchange, with $f_V$
representing the fraction of vector exchange in the interaction. The
short distance potential is \cite{grr1}
\begin{equation}\label{VS}
V_S=V_{HF}+V_{LS}+V_T+V_{SI}+V_{MIX},
\end{equation}
with
\begin{subequations}
\label{allpot}
\begin{eqnarray}
V_{HF}&=&\frac{32\pi\al_S\vec{S}_1\Dot\vec{S}_2}{9m_1m_2}\left\{(1-\frac{19\al_S}{6\pi})\de(\vec{r})-
\frac{\al_S}{8\pi}(8\frac{m_1-m_2}{m_1+m_2}+\frac{m_1+m_2}{m_1-m_2}\ln\frac{m_2}{m_1})\de(\vec{r})\right. \nonumber \\
& &\left.-\frac{\al_S}{24\pi^2}(33-2n_f)\nabla^2\left[\frac{\ln\,\mu r+\ga_E}{r}\right]+\frac{21\al_S}{16\pi^2}\nabla^2\left[\frac{\ln\, (m_1m_2)^\frac{1}{2}r+\ga_E}{r}\right]\right\}\label{pota} \\
V_{LS}&=&\frac{\al_S\vec{L}\Dot\vec{S}}{3m_1^2m_2^2r^3}\left\{\left[(m_1+m_2)^2+2m_1m_2\right]\left[1-\frac{3\al_S}{2\pi}+
\frac{\al_S}{6\pi}(33-2n_f)\left(\ln\mu r+\ga_E-1\right)\right]\right. \nonumber \\
& & \left.+\frac{\al_s}{2\pi}(m_1+m_2)^2\left[\frac{8}{3}-6(\ln\, (m_1m_2)^\frac{1}{2}r+\ga_E-1)\right]-\frac{3\al_s}{2\pi}(m_1^2-m_2^2)\ln\frac{m_2}{m_1}\right\}\label{potb}\\
[4pt] V_{T\;}
&=&\frac{4\al_S(3\vec{S_1}\Dot\hat{r}\vec{S_2}\Dot\hat{r}-\vec{S_1}\Dot\vec{S_2})}
{3m_1m_2r^3}\left\{1+\frac{4\al_S}{3\pi}+\frac{\al_S}{6\pi}\left[(33-2n_f)\left(\ln\mu
r+\ga_E-\frac{4}{3}\right)\right.\right. \nonumber \\
& &\left.\left.-18\left(\ln (m_1m_2)^\frac{1}{2}r+\ga_E-\frac{4}{3}\right)\right]\right\}\label{potc} \\
V_{SI}&=&\frac{2\pi\al_S}{3}(\frac{1}{m_1^2}+\frac{1}{m_2^2})\left\{(1-\frac{3\al_S}{2\pi})
\de(\vec{r})-\frac{\al_S}{24\pi^2}(33-2n_f)\nabla^2\left[\frac{\ln\,\mu
r+\ga_E}{r}\right]\right. \nonumber \\
& & \left.-\frac{\al_S}{6\pi
r^2}\left[\frac{9(m_1+m_2)^2-8m_1m_2}{m_1m_2(m_1+m_2)}\right]\right\}
\label{potd}\\\nonumber
V_{MIX}&=&-\frac{\al_S\vec{L}\Dot(\vec{S_1}-\vec{S_2})}{3m_1^2m_2^2r^3}\left\{(m_1^2-m_2^2)\left[1-\frac{\al_S}{6\pi}+
\frac{\al_S}{6\pi}(33-2n_f)\left(\ln\mu
r+\ga_E-1\right)\right.\right. \nonumber \\
& & \left.\left.-\frac{3\al_S}{\pi}\left(\ln
(m_1m_2)^\frac{1}{2}r+\ga_E-1\right)\right]-\frac{3\al_s}{2\pi}(m_1+m_2)^2\ln
\frac{m_2}{m_1}\right\}\label{pote}\\\nonumber
\end{eqnarray}
\end{subequations}

We have chosen $H_0$ such that it contains the relativistic kinetic
energy and the leading order spin-independent portions of the
long-range confining potential and the one-loop QCD short-range
potential. It is important to recall that the potential given by
Eq.\,(\ref{VS}) does not reduce to the potential in Ref.1, due to
the presence of annihilation terms in the equal-mass quark-antiquark
potential. It should also be noted that in calculating the matrix
elements of the $\delta(\vec{r})$ terms in Eqs. (5a) and (5d), we
`soften' their singularity by adopting the quasistatic approximation
of Ref.\cite{gjrs}, which leads to the replacement
\begin{equation}
\de(\vec{r})\to\frac{m^2}{\pi r}e^{-2mr}\,
\end{equation}
where $m$ is the quark mass. This softening helps the stability of
the eigenvalue calculation.

\section{Calculational approach \label{III}}
The $c\bar{s}$ mass spectrum and corresponding wave functions are
obtained using the variational approach described in Ref.1. In this
approach, we expand the wave functions as
\begin{equation}\label{wavefun}
\psi^m_{j\ell s}(\vec{r})=\sum_{k=0}^n C_k\left(\frac{r}{R}\right)^{k+\ell} \!e^{-r/R} \mathcal{Y}^m_{j\ell s}(\Omega)\,,
\end{equation}
where $\mathcal{Y}^m_{j\ell s}(\Omega)$ denotes the orbital-spin
wave function for a specific total angular momentum $j$, orbital
angular momentum $\ell$ and total spin $s$. The $C_k$'s are
determined by minimizing
\begin{equation}E=\frac{\langle\psi\,|H\,|\psi\rangle} {\langle\psi\,|\psi\rangle}
\end{equation}
with respect to variations in these coefficients. This procedure
results in a linear eigenvalue equation for the $C_k$'s and the
energies, and is equivalent to solving the Schr\"odinger equation.
The wave functions corresponding to different eigenvalues are
orthogonal and the $k^{\rm th}$ eigenvalue $\lambda_k$ is an upper
bound on the exact energy $E_k$. For $n=14$, the lowest four
eigenvalues for any $\ell$ are stable to one part in $10^6$. We performed a perturbative calculation, using $H_0$ as the unperturbed Hamiltonian and all other terms treated as first-order perturbations.

An optimal set of the parameters $\alpha=(\alpha_1,\alpha_2,\cdots,\alpha_n)$ were found by minimizing the $\chi^2$ function
\begin{equation}
\chi^2 = \sum_{i=1}^N\frac{\left(\mathcal{O}_{\rm exp\,i}-\mathcal{O}_{\rm th}(\alpha)_i\right)^2}{\sigma^2_i}\,,
\end{equation} 
where the $\mathcal{O}_i$ denote the experimental and theoretical values of some quarkonium observable and the $\sigma_i$ are the associated errors. In this work, the $\mathcal{O}_{\rm exp\,i}$ consist of a subset of the measured $D_s$ masses. For the masses, the $\sigma_i$ are taken to be the actual experimental error and a common intrinsic theoretical error added in quadrature. The latter error reflects the theory uncertainty associated with omitting corrections beyond one-loop and is estimated by requiring the $\chi^2$/degree of freedom to be approximately unity. Typically, this error is a few MeV. The minimization of $\chi^2$ with respect to  variations of the parameters $\alpha$ is accomplished using the search program STEPIT \cite{step}. The choices of $\al_S$ and $m_c$ were kept consistent with the results of running these parameters from the charmonium scale of Ref.\cite{rr1} by introducing a Gaussian prior in the $\chi^2$ function for each one.  For additional discussion of calculational details, see Appendix A of Ref.1.

\section{RESULTS \label{IV}}

We summarize our results in the following tables.  The parameters resulting from our fit are given in Table \ref{params}.

\begin{table}[h]
\centering
\begin{tabular}{l|d}
\toprule
\Sc$A$ (GeV$^2$) & 0.115 \\
\hline
\Sc $\al_S$      & 0.391 \\
\hline
\Sc $m_c$ (GeV)  & 1.66  \\
\hline
\Sc $m_S$ (GeV)  & 0.346  \\
\hline
\Sc $\mu$ (GeV)  & 1.19  \\
\hline
\Sc $f_V$        & 0.00  \\
\botrule
\end{tabular}
\caption{Fitted Parameters for the $c\bar{s}$ system\label{params}}
\end{table}
The results for our determination of the $D_s$ levels are shown in
Table \ref{dsspec} \cite{pdg}.
\begin{table}[h]\centering
\begin{tabular}{l|d|d} \toprule
\multicolumn{1}{c}{\Sc}$m_{c\bar{s}}$\,(MeV)\Sw  &\multicolumn{1}{c}{Model}  & \multicolumn{1}{c}{ Expt} \\
\hline
\Sb$D_s$               & 1968.49   & 1968.49\pm 0.34    \\
\hline
\Sb$D_s^*$             & 2112.3   & 2112.3\pm 0.5    \\
\hline
\Sb$D_{s0}(2317)$      & 2317.8   & 2317.8\pm 0.6    \\
\hline
\Sb$D_{s1}(2460)$      & 2459.6   & 2459.6\pm 0.6    \\
\hline
\Sb$D'_{s1}(2536)$     & 2535.2   & 2535.35\pm 0.34  \\
\hline
\Sb$D_{s2}(2573)$      & 2572.9   & 2572.6\pm 0.9    \\
\hline
\Sb$D_s(2486)$         & 2485.8   &                  \\
\hline
\Sb$D_s^*(2637)$       & 2637.2   & 2690.\pm 7        \\
\botrule
\end{tabular}
\caption{Results for the $D_s$ spectrum are shown. The fit uses all
the indicated states of the $D_s$ system except for the $n=2$ $D_s$ and $D^*_s$.} \label{dsspec}
\end{table} 
Overall our fit to the spectrum is quite good.

As is usual in potential model treatments \cite{grr2,god,beh,elq1,bgs}, the radiative widths were calculated in the dipole approximation. We obtained the $E_1$ matrix elements by using the variational radial wave functions to construct initial and final state wave functions with the appropriate angular dependence and explicitly performing the angular integration.  Our results are
\begin{equation}
\Gamma_{\!fi}=\frac{4\al}{9}\left(\frac{q_1m_2-q_2m_1}{m_1+m_2}\right)^2\omega^3
|\langle f|r|i\rangle|^2\frac{E_f}{M_i}\left\{
\begin{array}{ccc}
1          & {\rm for} & ^3P_J\to ^3\!\!S_1  \\[4pt]
1          & {\rm for} & ^1P_1\to ^1\!\!S_0  \\[4pt]
(2J+1)/3   & {\rm for} & ^3S_1\to ^3\!\!P_J  \\[4pt] 
3          & {\rm for} & ^1S_0\to ^1\!\!P_1  
\end{array} \right. 
\end{equation}
for $E_1$ transitions. Here, $\omega$ is the photon energy, $q_1$
and $q_2$ are the quark charges in units of the proton charge, $E_f$
is the energy of the final quarkonium state, $M_i$ is the mass of
the initial quarkonium state, and $m_1$ and $m_2$ are the quark
masses.

We also take into account the mixing between the $^1P_1$ and $^3P_1$ eigenstates of the $c\bar{s}$ Hamiltonian due to the $\vec{L}\Dot(\vec{S}_1-\vec{S}_2)$ terms in Eqs.\,(\ref{VL}) and (\ref{pote}) of the perturbative potential. This mixing yields the two $J=1$ states $D_{s\,1}$ and $D'_{s\,1}$. They are, explicitly,
\begin{subequations}\label{mix}
\begin{eqnarray}
|D_{s\,1}(2460)\rangle&=&\sin(\th)|^3P_1\rangle+\cos(\th)|^1P_1\rangle \,, \label{minus} \\[4pt]
|D'_{s\,1}(2536)\rangle&=&\cos(\th)|^3P_1\rangle-\sin(\th)|^1P_1\rangle \,, \label{plus}
\end{eqnarray}
\end{subequations}
where
\begin{equation}\label{tan}
\tan(\th)=-\frac{V_{31}}{E_+-E(^1P_1)}\,,
\end{equation}
with $V_{31}$ denoting the expectation value of the mixing terms and $E_+$ denoting the larger of the two eigenvalues of the mixing matrix. Note that, because of the $1/m^2_2$ behavior of the these terms, the mixing is quite sensitive to the strange quark mass.  The conventions used in parameterizing the mixing are given in the Appendix.

For $M_1$ transitions, a parallel calculation, using the fact that the singlet and triplet $s-$states have the same wave functions in the
perturbative treatment, yields

\begin{equation}
\Gamma_{\!fi}=\frac{4\al}{3}\omega^3
\left(\frac{q_1}{2m_1}-\frac{q_2}{2m_2}\right)^2\frac{E_f}{M_i}
\end{equation}
for the $D^*_s\to D_s+\ga$. In the case of the $p-$state magnetic transitions $D_{s\,1}\to D_{s\,0}+\ga$ and $D'_{s\,1}\to D_{s\,0}+\ga$, both the singlet and triplet states of the mixtures in Eq.\,(\ref{mix}) contribute to the widths. If we use the perturbative wave functions, then the relative phase of the triplet contribution with respect to the singlet contribution is $\pi/2$. The widths in this case are
\begin{equation}
\Gamma_{\!fi}=\frac{4\al}{9}\omega^3\frac{E_f}{M_i}
\left\{\begin{array}{ccr}
2\sin^2(\th)\left(\ds \frac{q_1}{2m_1}+\frac{q_2}{2m_2}\right)^2+\cos^2(\th) \left(\ds\frac{q_1}{2m_1}-\frac{q_2}{2m_2}\right)^2 & {\rm for} & D_{s\,1}\to D_{s\,0}+\ga     \\ [4pt]
2\cos^2(\th)\left(\ds\frac{q_1}{2m_1}+\frac{q_2}{2m_2}\right)^2+\sin^2(\th) \left(\ds\frac{q_1}{2m_1}-\frac{q_2}{2m_2}\right)^2  & {\rm for} & D'_{s\,1}\to D_{s\,0}+\ga
\end{array}\right.
\end{equation}
The resulting radiative widths are shown in Table \ref{onegam}.
\begin{table}[h]
\centering
\begin{tabular}{l|r|r} \toprule
\multicolumn{1}{c}{\Sc $\Gamma_{\ga}$\,(keV)}  &\multicolumn{1}{c}{Model}  &\multicolumn{1}{c}{ Expt}   \\
\hline
\Sb$D_s^*\to D_s$\mbox{\rule{12pt}{0pt}}& 1.91      & $< 1.9\times 10^3$ \\
\hline
\Sb$D_{s\,0}(2317)\to D_s^*$            & 4.92        &                   \\
\hline
\Sb$D_{s\,1}(2460)\to D_s$              & 12.8        & BR =$0.18\pm 0.04$\\
\hline
\Sb$D_{s\,1}(2460)\to D_s^*$            & 15.5        & BR$< 0.08$        \\
\hline
\Sb$D_{s\,1}(2460)\to D_{s\,0}(2317)$   & 5.74        &                   \\
\hline
\Sb$D'_{s\,1}(2536)\to D_s$             & 54.5        &                   \\
\hline
\Sb$D'_{s\,1}(2536)\to D_s^*$           & 8.90        & {\rm possibly seen}     \\
\hline
\Sb$D'_{s\,1}(2536)\to D_{s\,0}(2317)$  & 2.36        &                   \\
\hline
\Sb$D_{s\,2}(2575)\to D_s^*$            & 44.1        &                   \\
\hline
\Sb$D^*_s(2637)\to D_{s\,0}(2317)$      & 6.76        &                   \\
\hline
\Sb$D^*_s(2637)\to D_{s\,1}(2460)$      & 2.8         &                   \\
\hline
\Sb$D^*_s(2637)\to D'_{s\,1}(2536)$     & 0.24        &                   \\
\hline
\Sb$D^*_s(2637)\to D_{s\,2}(2573)$      & 0.35        &                   \\
\hline
\Sb$D_s(2485)\to D_{s\,1}(2460)$        & 0.01        &                   \\
\botrule
\end{tabular}
\caption{The the radiative decays of the $D_s$ mesons are shown. These widths are computed using the mass values obtained directly from our calculation. This includes the $n=2$ pseudoscalar and vector states, the latter of which has recently been observed with a higher mass \cite{pdg}.  The widths are from \cite{pdg}. \label{onegam}}
\end{table}

\section{CONCLUSIONS \label{V}}

We have shown that a potential model consisting of the relativistic kinetic energy, a linear long-range confining potential together with its $v^2/c^2$ relativistic corrections, and the full $v^2/c^2$ plus one-loop QCD corrected short distance potential is capable of providing extremely good fits to the spectra of the $D_s$ states by treating them as states of the $c\bar{s}$ system. We find that in this perturbative treatment the long-range potential must be entirely due to scalar exchange.

The single photon widths can be obtained from the variational wave functions, but, apart from some branching ratio measurements, there are relatively little data available. Our theoretical results are comparable to those given in Refs.\cite{god} and \cite{beh} allowing for the fact that both of these references use a substantially higher strange quark mass (419 MeV and 480 MeV, respectively). In every case, efforts to model these states will be greatly improved by the availability of additional data.

\begin{acknowledgments}
We would like to thank Christopher Potter for valuable discussions
and assistance with some computational issues. SFR would like to
thank the Kavli Institute for Theoretical Physics for its
hospitality during July, 2008. This research was supported
in part by the National Science Foundation under Grants PHY-0555544 and PHY-0551164.
\end{acknowledgments}

\appendix
\section{Details of mixing}
The mixing of the $^3P_1$ and $^1P_1$ states is obtained by diagonalizing the $2\times 2$ matrix
\begin{equation}
\left(\begin{array}{cc}
E_3       & V_{31} \\
V_{31}    & E_1
\end{array}\right)\,,
\end{equation}
where $E_3$ is the $^3P_1$ energy, $E_1$ is the $^1P_1$ energy and $V_{31}$ is the mixing matrix element. In perturbation theory, this is relatively simple since all of these matrix elements can be calculated using the unperturbed wave functions that are all the same. The energy eigenvalues are
\begin{equation}
E_\pm = \frac{1}{2}\left(E_3+E_1\right)\pm\frac{1}{2} \sqrt{(E_3-E_1)^2+4V_{31}^2}\,,
\end{equation}
and we fit the $D'_{s1}(2536)$ and $D_{s1}(2460)$ to $E_+$ and $E_-$. Note that as $V_{31}\to 0$, $E_+\to E_3$ and $E_-\to E_1$. To define the mixing angles in terms of known parameters, we assume that the eigenvector $\psi_+$ corresponding to $E_+$ behaves as 
\begin{equation}
\psi_+ \stackrel{V_{31}\to 0}\longrightarrow \left(\begin{array}{c}
                                                   1  \\
                                                   0
                                                   \end{array}\right)\,.
\end{equation}
With this assumption, $\psi_+$ is
\begin{equation}
\psi_+=\frac{1}{\sqrt{\ds (E_+-E_1)^2+V_{31}^2}}\left(\begin{array}{c}
                                                      E_+ - E_1   \\
                                                      V_{31}
                                                      \end{array}\right)
\end{equation}
By writing $\psi_+$ as
\begin{equation}
\psi_+=\left(\begin{array}{c}
             \cos(\th)  \\
             -\sin(\th)
             \end{array}\right)\,,
\end{equation}
with
\begin{equation}
\tan(\th)=-\frac{V_{31}}{E_+-E_1}\,,
\end{equation}
we arrive at the decomposition Eq.\,(\ref{mix}). It is possible to obtain an estimate of the mixing angle by using the branching ratios of the $D_{s1}(2460)\to D_s\,\ga$ and $D_{s1}(2460)\to D^*_s\,\ga$, whose ratio gives
\begin{equation}
\frac{\Ga(D_{s1}(2460)\to D^*_s\,\ga)\,\om^3}{\Ga(D_{s1}(2460)\to D_s\,\ga)\,\om^{*\,3}} =\tan^2(\th)\,,
\end{equation}
where $\om$ is the momentum of the photon in the $D_s$ transition and $\om^*$ is the corresponding photon momentum in the $D^*_s$ transition. Using the published branching ratio information, with favorable assumptions, the data are consistent with $\th=\pm 50^{\,\circ}$. Our calculation gives $\th=59.7^{\,\circ}$.

\end{document}